\documentclass[11pt,twoside,a4paper,cmspaper,final,collab]{cms-tdr}
\def\svnVersion{8c8104f}\def\svnDate{2026/11/04}\def\cmsCernNoTag{CERN-EP-2026-144}\def\cmsCernDate{\today}\def\cmsMessage{Published in Physical Review Letters as \href{http://dx.doi.org/10.1103/jydv-n2js}{\doi{10.1103/jydv-n2js}.}}
\begin{document}\cmsNoteHeader{SMP-23-001}

\newcommand{\WW}{\ensuremath{\PW\PW}\xspace}
\newcommand{\ZZ}{\ensuremath{\PZ\PZ}\xspace}
\newcommand{\WZ}{\ensuremath{\PW\PZ}\xspace}
\newcommand{\detajj}{\ensuremath{\abs{\Delta\eta_{\mathrm{jj}}}}\xspace}
\newcommand{\mjj}{\ensuremath{m_{\mathrm{jj}}}\xspace}
\newcommand{\DGNN}{D_{\mathrm{GNN}}^{\text{flat}}\xspace}
\providecommand{\PGgst}{\HepParticle{\PGg}{}{\ast}\xspace}
\newlength\cmsTabSkip\setlength{\cmsTabSkip}{1ex}
\newlength{\cmsFigWidth}
\ifthenelse{\boolean{cms@external}}
{\setlength{\cmsFigWidth}{\columnwidth}
\providecommand{\cmsLeft}{upper\xspace}
\providecommand{\cmsRight}{lower\xspace}}
{\setlength{\cmsFigWidth}{0.6\textwidth}
\providecommand{\cmsLeft}{left\xspace}
\providecommand{\cmsRight}{right\xspace}}

\ifthenelse{\boolean{cms@external}}{
    \newcommand{\appname}{End Matter}
    \def\appendixname{\appname}
    \providecommand{\EndMatter}{ in the \appname\xspace}
}{
\providecommand{\EndMatter}{ in the \nameref{endmatter}\xspace}
}
\cmsNoteHeader{SMP-23-001}
\title{Observation of electroweak production of pairs of \texorpdfstring{\PZ}{Z} bosons in proton-proton collisions at 13\TeV}

\date{\today}

\abstract{The first evidence of electroweak (EW) production of pairs of \PZ bosons in association with two jets (\text{jj}) in the final state $\PZ\PZ \text{jj}\to\Pell\Pell\PGn\PGn \text{jj}$, where $\Pell = \Pe, \PGm$, is reported by the CMS experiment. The analysis is based on a data sample of proton-proton ($\Pp\Pp$) collisions at $\sqrt{s} = 13\TeV$, corresponding to an integrated luminosity of 138\fbinv.  Events are selected by requiring exactly two leptons of same flavor and opposite charge, large missing transverse momentum, and two jets with a large rapidity separation and large invariant mass.   The EW production cross section in a fiducial volume is $\sigma_{\mathrm{EW}}(\Pp\Pp \to \PZ\PZ \text{jj}\to\Pell\Pell\PGn\PGn \text{jj}) = 0.37^{+0.14}_{-0.12}\stat^{+0.06}_{-0.06}\syst$\unit{fb},  in agreement with the standard model prediction of $0.39 \pm 0.06$\unit{fb}. The observed (expected) significance of the signal is 3.1 (2.8) standard deviations.  Limits on anomalous quartic gauge couplings are derived in terms of dimension-8 effective field theory operators.  A combination with the previously reported result from the \ZZ decay channel with four charged leptons yields an observed (expected) significance of 5.0 (4.5) standard deviations for the EW production of \PZ boson pairs.}

\hypersetup{
pdfauthor={CMS Collaboration},
pdftitle={Observation of electroweak production of pairs of Z bosons in proton-proton collisions at 13 TeV with the CMS detector},
pdfsubject={CMS},
pdfkeywords={CMS, electroweak, ZZ, Run2}}

\maketitle 

In the standard model (SM) of particle physics, the \PW and \PZ bosons acquire their masses via the Brout--Englert--Higgs mechanism. A direct consequence of this mechanism is that Higgs boson exchange plays an essential role in vector boson scattering (VBS) processes---the electroweak (EW) production of gauge boson pairs---by ensuring that scattering amplitudes do not exceed the unitarity limit at high energies~\cite{Lee:1977eg}.
In VBS processes, vector bosons radiated from initial-state quarks interact to produce new boson pairs.  This analysis focuses on the VBS process with the EW production of two \PZ bosons accompanied by two jets, using proton-proton ($\Pp\Pp$) collision data collected with the CMS detector at the CERN LHC with $\sqrt{s}=13\TeV$ during 2016--2018, corresponding to an integrated luminosity of 138\fbinv. We investigate scenarios where one \PZ boson decays into charged leptons (\Pe or \PGm), and the other into a neutrino pair. These results are combined with the results previously reported in the channel with four charged leptons~\cite{CMS:2020fqz} to determine the statistical significance of the measurement of the EW \ZZ production.

\begin{figure}[htbp]
    \centering
    \includegraphics[page=1,width=0.5\textwidth]{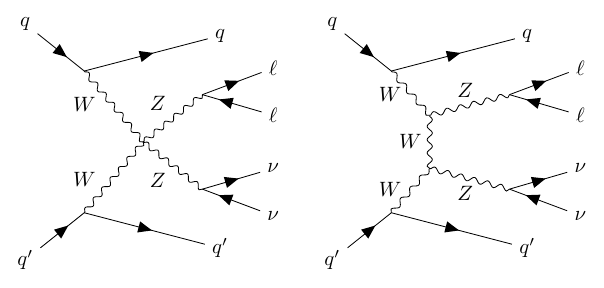}
    \includegraphics[page=2,width=0.5\textwidth]{Figure_001.pdf}
    \caption{Example Feynman diagrams of the signal EW \ZZ process (first three diagrams) and the QCD \ZZ ($\PQq\PAQq$) background process (last diagram).}
    \label{fig:feyn}
\end{figure}

In the analysis presented here, we select candidate events with exactly two identified leptons (electrons or muons) of opposite sign and same flavor (OSSF), large missing transverse momentum (\ptmiss)~\cite{CMS:2019ctu}, and two jets with a large rapidity separation and large invariant mass.
Figure~\ref{fig:feyn} shows examples of Feynman diagrams for the $\mathcal{O}(\alpha_{\mathrm{EW}}^6)$ VBS processes (EW \ZZ), as well as the $\mathcal{O}(\alpha_{\mathrm{EW}}^4\alpS^2)$ quantum chromodynamics (QCD) induced process (QCD \ZZ) that can produce the same final state and represents the irreducible background, where $\alpha_{\mathrm{EW}}$ and $\alpS$ are the electroweak and strong couplings, respectively. The other important backgrounds are \WZ and \WW production with leptonic decays of the \PW and \PZ bosons, \ttbar, and Drell--Yan (DY) production of $\PZ + \text{jets}$.

Besides measuring the EW cross section, we use this data sample to probe anomalous quartic gauge couplings (aQGCs)~\cite{Almeida:2020bzy},  which correspond to beyond-the-SM (BSM) interactions among four gauge bosons. Such effects can modify the production of \ZZ pairs in VBS processes.
In this analysis, possible deviations from the SM are parameterized using dimension-8 operators in the SM effective field theory (SMEFT) framework~\cite{Anders:2018oin,Eboli:2006wa}.

The CMS Collaboration has observed the EW 
production of same-sign \WW~\cite{CMS:2017fhs}, opposite-sign \WW~\cite{CMS:2022woe}, and 
\WZ~\cite{CMS:2020ssj} and previously obtained evidence for 
EW \ZZ production in the four-lepton ($\Pell\Pell\Pellpr\Pellpr$) final state with a significance of 4.0 standard deviations~\cite{CMS:2020fqz}. The ATLAS Collaboration observed EW \ZZ production using the combination of $\Pell\Pell\Pellpr\Pellpr$ and $\Pell\Pell\PGn\PGn$ final states, with a significance of 5.7 standard deviations~\cite{ATLAS:ZZ_VBS}. 

The CMS apparatus~\cite{CMS:2008xjf,CMS:2023gfb} is a multipurpose, nearly hermetic detector, designed to trigger on~\cite{CMS:2020cmk,CMS:2016ngn,CMS:2024aqx} and identify electrons, muons, photons, and (charged and neutral) hadrons~\cite{CMS:2020uim,CMS:2018rym,CMS:2014pgm}. A global ``particle-flow'' (PF) algorithm~\cite{CMS:2017yfk} aims to reconstruct all individual particles in an event, combining information provided by the all-silicon tracker and by the crystal electromagnetic and brass-scintillator hadron calorimeters, operating inside a 3.8\unit{T} superconducting solenoid, with data from the gas-ionization muon detectors interleaved with the layers of the flux-return yoke outside the solenoid. The reconstructed particles are used to build \PGt leptons, jets, and \ptmiss~\cite{CMS:2018jrd,CMS:2016lmd,CMS:2019ctu}. 

Jets are reconstructed by clustering PF candidates using the anti-\kt algorithm~\cite{Cacciari:2008gp,Cacciari:2011ma} with a distance parameter of 0.4. 
Jets originating from the hadronization of \PQb quarks are identified using the \textsc{DeepJet} algorithm~\cite{BTV-16-002,Bols:2020bkb,CMS-DP-2018-058} based on properties of secondary vertices and track impact parameters.
In this analysis, \PQb tagging is used to reject events with genuine \PQb quark jets in the final state. 
The missing transverse momentum vector \ptvecmiss is computed as the negative vector sum of the transverse momenta of all the PF candidates in an event.

We model signal and background events with Monte Carlo (MC) simulations, reweighted to account for known discrepancies between data 
and simulated events.  The efficiencies for electron and muon reconstruction, identification, and isolation as functions of the transverse momentum (\pt) and pseudorapidity ($\eta$) of lepton, are extracted from events 
with leptonic \PZ boson decays using a ``tag-and-probe'' technique~\cite{CMS:2011aa}.
For each data set, simulated events are reweighted according to the pileup (additional $\Pp\Pp$ inelastic interactions in the same or adjacent bunch crossings) profile distribution observed in data.
Some of the background estimates, notably the DY production, are derived using control samples in data.
The EW signal process $\Pp\Pp\to\PZ\PZ\PQq\PAQq$ is simulated at next-to-leading order (NLO) in QCD with \POWHEG v2~\cite{Alioli:2010xd} interfaced with the event generator \PYTHIA 8.240~\cite{Sjostrand:2014zea} for parton showering and hadronization.  
The dipole approach~\cite{Cabouat:2017rzi} is used to model the initial-state radiation, rather than the standard \pt-ordered one used in the \PYTHIA 
parton shower simulation, as the latter does not properly describe extra QCD emissions in the vector boson fusion and VBS processes~\cite{Jager:2020hkz}.
The QCD-induced \WW and \ZZ backgrounds from quark-antiquark ($\PQq\PAQq$) initial states are also modeled with \POWHEG v2. In this setup, additional jet emissions beyond the first are generated by the parton shower and are described at leading order accuracy.
The size of the interference between the EW- and QCD-induced processes was evaluated from simulation to be 8\% of the EW signal yield after signal region (SR) selection, and we treat this as a systematic uncertainty.
We model \WW and \ZZ boson production induced by gluons (gg) with \MCFM7.0~\cite{MCFM}, while \POWHEG v2 is used to simulate \ttbar~\cite{Frixione:2007nw} and single top quark production in association with a \PW boson~\cite{Re:2010bp}, interfaced with PYTHIA for parton showering and hadronization.
The default parton distribution function (PDF) set NNPDF3.1~\cite{NNPDF:2017mvq} at next-to-NLO (NNLO) accuracy, and the underlying event tune 
CP5~\cite{CMS:2019csb} are common to all simulated events.
All generated events are processed through a simulation of the CMS detector based on \GEANTfour~\cite{GEANT4:2002zbu} 
and are reconstructed with the same algorithms as used for data.

Events are selected using a combination of single- and double-lepton triggers requiring isolated electrons or muons. Exactly two prompt electrons or muons are required, with $\pt > 25$ (20)\GeV for the leading (subleading) lepton.  
In the SR, a pair of OSSF leptons is required, and events with additional leptons with $\pt > 10\GeV$ are vetoed. The dilepton system is required to have $\pt > 60\GeV$ and an invariant mass between 76 and 106\GeV, consistent with the decay products of a \PZ boson.
We require $\ptmiss>120\GeV$ to reduce the DY background. A veto is applied to reject events containing hadronically decaying tau leptons to reduce \WZ background.
We also veto events containing \PQb jet candidates with $\pt>30\GeV$ to reduce contributions from top quark processes. 
To enhance the relative contribution from VBS processes, we select events with two or more jets and require the two highest \pt (leading) jets to have a dijet mass $\mjj>400\GeV$ and a large $\eta$ separation $\detajj>2.5$.

In addition, two control regions (CRs) orthogonal to the SR are defined to help assess the background contributions. 
The $3\Pell$ CR, which is dominated by \WZ events, requires exactly one additional lepton with $\pt>10\GeV$,
in addition to all the SR requirements. The \pt of the third lepton is summed vectorially with \ptmiss to emulate a $\PZ+\ptmiss$ signal-like topology.
The $\Pe\PGm$ CR, dominated by contributions from $\PW^+\PW^-$ and \ttbar production, is otherwise identical to the SR except that the two leptons must be one electron and one muon. In both CRs, $\ptmiss>70\GeV$ is required.

We use input from both simulation and data to determine the background contributions.  
For processes that produce final states that contain neutrinos, including \WW, \WZ,  and \ttbar production, we rely on simulation to represent the kinematic characteristics of the events, while their normalizations are constrained using the dedicated CRs. 
The distributions from both the $3\Pell$ and $\Pe\PGm$ CRs are included in a maximum likelihood fit to the data in the SR and background CRs, taking into account correlations among the systematic uncertainties.
For the DY background, where the apparent \ptmiss results from mismeasurement,  both the normalization and the dilepton \ptmiss and \pt distributions are estimated from a photon-plus-jets data control sample, as described below.

The DY process is a significant reducible background with no genuine \ptmiss.
When detector instrumental effects result in large apparent \ptmiss, these DY events can enter the SR. 
Despite suppression by the requirement $\ptmiss > 120\GeV$, the large cross section of DY production requires us to estimate this background carefully. 
We employ a control sample of single-photon events ($\PGg+\text{jets}$) in data as a proxy for $\PZ + \text{jets}$ events, as described\EndMatter.

To further enhance the separation of the rare EW \ZZ signal from backgrounds, we employ a multivariate discriminator based on a graph neural network (GNN)~\cite{Moreno:2019bmu}, which operates on a graph-structured representation of the data. In this framework, individual physics objects (\eg, leptons and jets) are represented as nodes, while pairwise relationships or potential correlations
between them are modeled as edges. The network is trained on simulated events using a supervised learning approach. 
Individual kinematic quantities are provided to the GNN so that it can learn the subtle differences between signal and background. These input variables include the \pt, $\eta$, and azimuthal angle $\phi$ of the charged leptons and the two leading jets, and also a third jet if present. The distributions of the input variables are well modeled by the simulation and have been studied in~\cite{CMS:2022woe,CMS:2022ley}. The GNN is trained using simulated samples corresponding to 2018 data-taking conditions, and the performance is stable across all years. Additional inputs correspond to global event variables: the number of jets, \ptmiss and its corresponding $\phi$, the $\mjj$ and $\detajj$ of the two leading jets.

A binned maximum likelihood fit is applied to the flattened distributions of the GNN classifier score ($\DGNN$), combining all data-taking periods, using the CMS \textsc{Combine} package~\cite{CMS:2024onh} to estimate the contributions from signal and background processes. All results presented for the SM signal measurement in this analysis are derived from this likelihood fit strategy. The flattening transformation is constructed from the weighted cumulative distribution of the signal GNN score, such that the signal distribution is mapped to a uniform distribution. The same transformation is applied to all other processes. 
Since electrons and muons have similar sensitivities, they are combined in the discriminating fit. 
For each individual bin, a Poisson likelihood term describes the fluctuation of the data around the expected central value, which is given by the sum of the contributions from signal and background processes, and the Barlow--Beeston ``lite'' method~\cite{BARLOW1993219,Conway:2011in} is used for the treatment of MC statistical uncertainties.

Systematic uncertainties are incorporated as nuisance parameters in the likelihood fit, 
accounting for both normalization and shape effects on the signal and background predictions. 
The impact of each source is evaluated by varying the corresponding parameters within their uncertainties and propagating the resulting variations through the full analysis.
The estimation of the contributions of individual sources of systematic uncertainty is described in more detail\EndMatter. The dominant systematic uncertainties arise from theoretical modeling, 
in particular renormalization and factorization scale variations and PDFs, 
as well as experimental effects related to the jet energy scale (JES) and resolution (JER). 
Lepton efficiencies, pileup modeling, and missing higher-order QCD--EW corrections contribute smaller uncertainties.

\begin{figure}
    \centering   
    \includegraphics[width=0.45\textwidth]{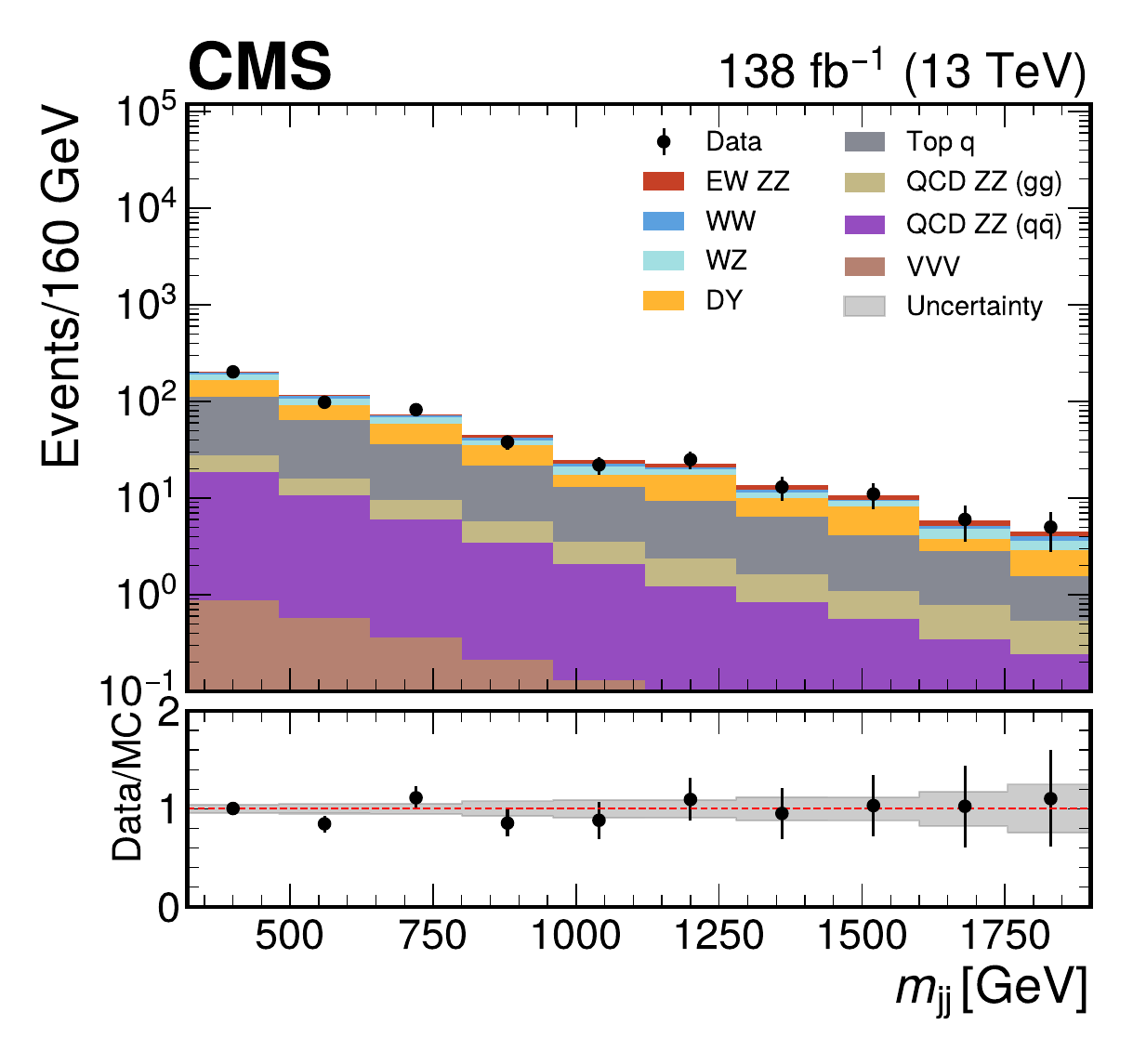}
    \includegraphics[width=0.45\textwidth]{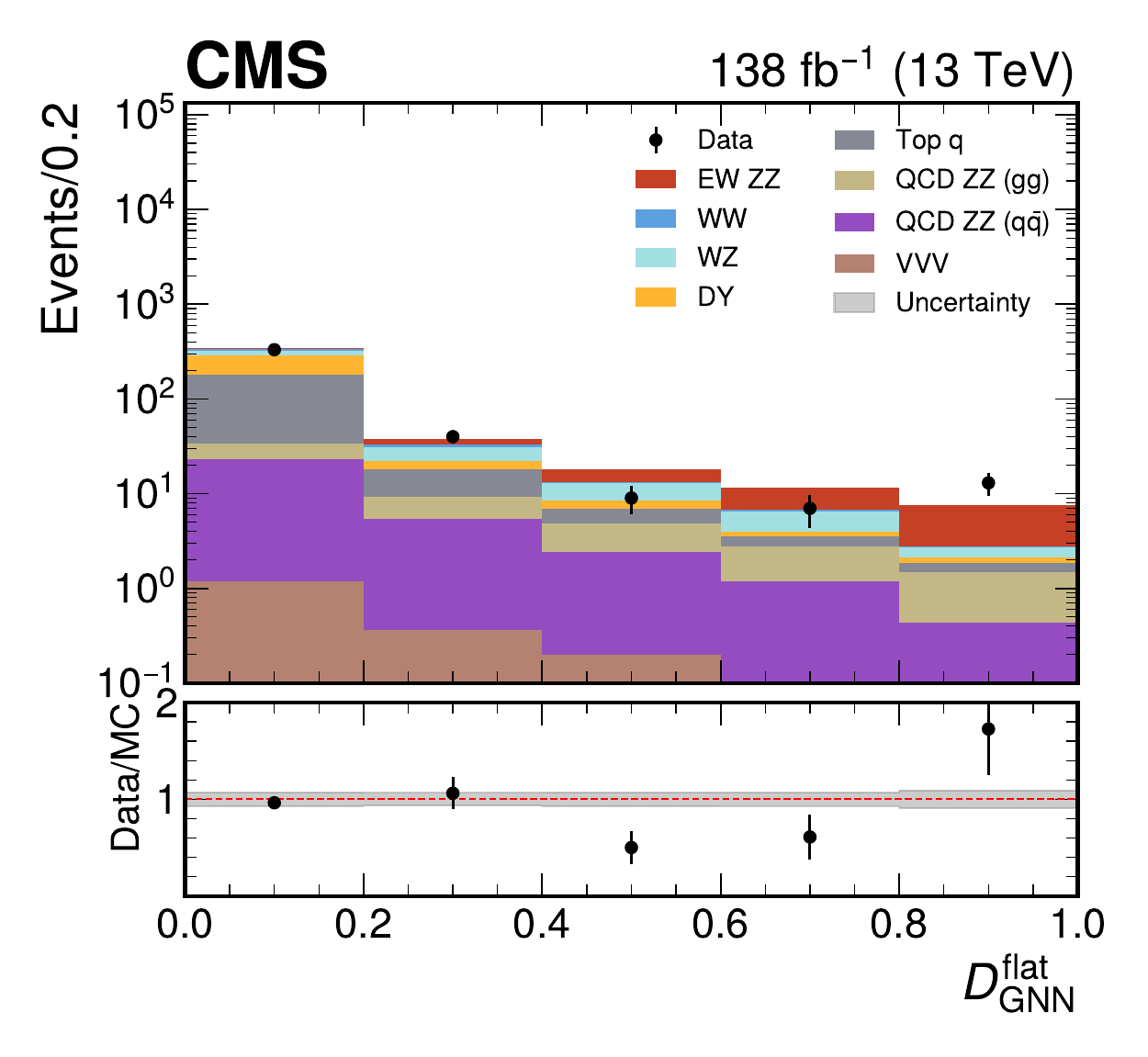}
    \caption{
    Distributions of the dijet invariant mass (\cmsLeft) and the flattened GNN discriminator (\cmsRight) in the SR after the maximum likelihood fit. 
    The data are shown as black points, whereas the stacked histograms represent the fitted signal and background contributions. The vertical bars on the data points represent the statistical uncertainties. The gray band indicates the total uncertainty in the prediction after the fit, including both statistical and systematic components.}
    \label{fig:prepostfit}
\end{figure}

The estimated yields obtained from the fit for the signal and all backgrounds are shown in Table~\ref{tab:SR_yields_summary}. The expected results are predicted using an Asimov data set~\cite{Cowan:2010js}, where central values of the templates are set in each bin to those provided by the prediction. The observed (expected) significance of the signal is 3.1 (2.8) standard deviations, evaluated using likelihood-based hypothesis tests~\cite{Cowan:2010js,CLS1,CLS2}. The $p$-value of the fit is 0.11. The $\mjj$ and $\DGNN$ distributions after the fit are shown in Fig.~\ref{fig:prepostfit}, with all 2016--2018 data sets combined. The binning of the $\DGNN$ distribution is determined from the expected significance evaluated using simulation. In the highest bins of the $\DGNN$ distribution, limited event counts lead to visible fluctuations in the Data-to-MC ratio.

\begin{table}[htbp]
    \centering
    \topcaption{Estimated event yields for signal and background processes in the SR after the maximum likelihood fit, with associated uncertainties including the statistical and systematic components. The QCD-induced production of \PZ boson pairs is shown separately for $\PQq\PAQq$ and $\Pg\Pg$ initial states. \PV denotes \PW or \PZ.}
    \label{tab:SR_yields_summary}
    \begin{scotch}{c c}
    Process & Yield (events) \\
    \hline
    DY & $111 \pm 8$ \\
    Top quark & $159 \pm 23$ \\
    $\PW\PW$ & $18 \pm 3$ \\
    $\PW\PZ$ & $55 \pm 3$ \\
    QCD $\PZ\PZ$ ($\PQq\PAQq$) & $31 \pm 2$ \\
    QCD $\PZ\PZ$ ($\Pg\Pg$) & $20 \pm 1$ \\
    VVV & $2 \pm 0.1$ \\
    Total background & $395 \pm 27$ \\
    Total signal (EW $\PZ\PZ$) & $22 \pm 2$ \\
    All processes & $416 \pm 26$ \\
    Data & $400$ \\
\end{scotch}
\end{table}

The same fit is used to measure the EW production cross section in a fiducial region defined by the following requirements on generator-level quantities.
Two \PZ bosons must come from EW production with one decaying to neutrinos and the other with $\pt>60\GeV$ decaying to electrons or muons. The \pt values of the leading and subleading leptons must be at least 25 and 20\GeV, respectively, with $\abs{\eta}<2.5$ for both leptons. The invariant mass of the dilepton system is required to be between 76 and 106\GeV. 
Leptons are defined as dressed leptons. The momenta of photons are added to a lepton if the separation $\Delta R = \sqrt{\smash[b]{(\Delta \eta)^2 + (\Delta \phi)^2}}<0.1$. Two generator-level jets, clustered from all stable particles except neutrinos, are required with $\pt>30\GeV$, $\abs{\eta}<4.7$, a separation $\detajj>2.5$ and $\mjj>400\GeV$. All leptons and jets are required to be separated from one another by at least $\Delta R=0.4$. The generator-level \ptmiss, defined as the vector sum of
the \pt of neutrinos, is required to satisfy $\ptmiss>120\GeV$. The QCD-induced contributions are estimated from simulation and are treated as background in the fit.
We find $\sigma_{\mathrm{EW}}^{\mathrm{fid}}(\Pp\Pp \to \ZZ\text{jj}\to\Pell\Pell\PGn\PGn\text{jj}) = 0.37^{+0.14}_{-0.12}\stat^{+0.06}_{-0.06}\syst$\unit{fb}
in agreement with the SM prediction of $0.39\pm 0.06$\unit{fb}. Based on SM simulation, we estimate that approximately 14\% of the accepted events pass the reconstruction-level selection but come from outside of the fiducial volume; these are treated as background in the fit.

\begin{figure}
    \centering
        \includegraphics[width=0.45\textwidth]{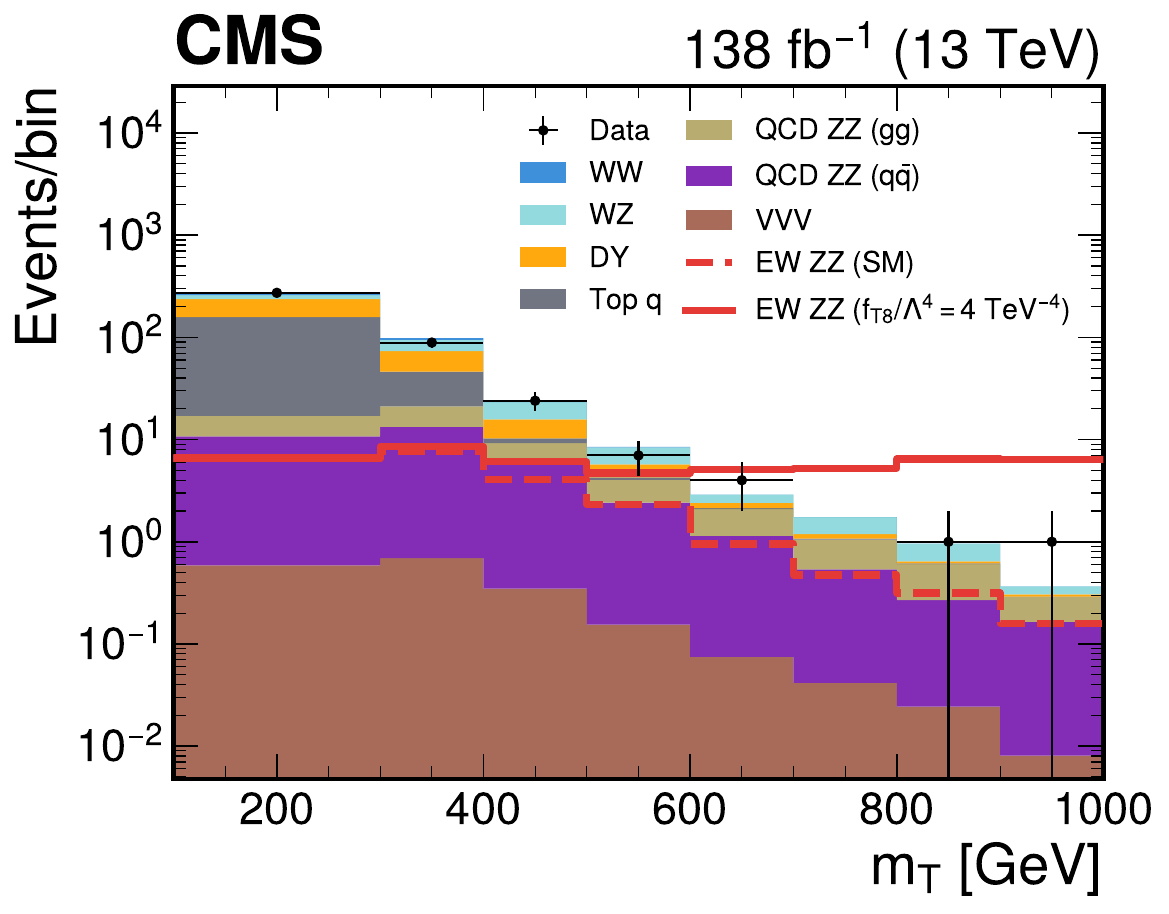}
        \includegraphics[width=0.45\textwidth]{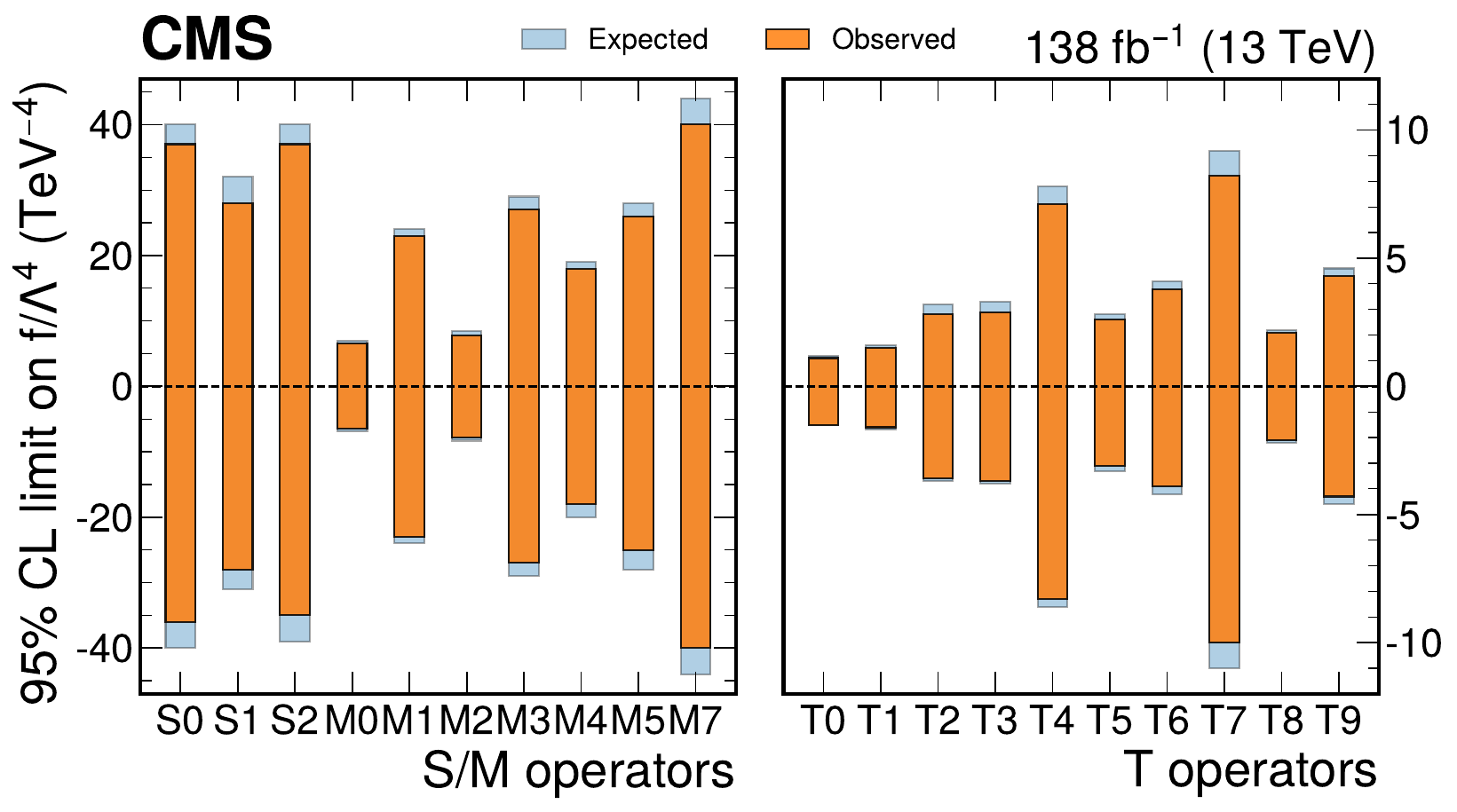}
    \caption{The $\mT$ distribution (\cmsLeft) with the expected contribution from a T8 operator, showing the SM case with all EFT Wilson coefficients set to zero (dashed red), and the case with $f_{\mathrm{T8}}/\Lambda^4 = 4\TeV{}^{-4}$ (solid red). The vertical bars on the data points represent the statistical uncertainties. The expected and observed limits (\cmsRight) at 95\% \CL are obtained from the likelihood fits.}
    \label{fig:EFT_limit}
\end{figure}

We study contributions of potential BSM physics scenarios in the $\Pell\Pell\PGn\PGn$ channel within the SMEFT framework~\cite{Anders:2018oin,Eboli:2006wa}. In general, this framework represents additional interactions beyond the SM with dimension-6 and higher operators, and we focus on a set of dimension-8 operators that characterize the four-boson interactions and contribution to potential aQGCs. The contributions from such operators tend to be enhanced at high \pt, and we selected the transverse mass
($\mT = \sqrt{2\smash[b]{ \pt^{\PZ} \ptmiss [1 - \cos{\Delta\phi(\vec{p}_{\text{T}}^{\PZ}, \ptvecmiss})]}}$) 
as a distribution that is sensitive to these contributions, as illustrated in Fig.~\ref{fig:EFT_limit} (\cmsLeft).
We fit using a maximum likelihood estimator to derive two-sided 95\% confidence
level (\CL) intervals on the couplings $f_i/\Lambda^4$ of dimension-8
operators describing anomalous quartic gauge interactions, where $f_i$ are dimensionless Wilson coefficients and $\Lambda$ denotes the energy scale of new physics in the effective field theory. Limits on each operator are obtained by varying one coefficient at a time,
while fixing all other coefficients to zero. The operators considered correspond to
$i=$ S0--S2, M0--M5 and M7, and T1--T9, representing the longitudinal (S), mixed (M),
and transverse (T) operator classes~\cite{Almeida:2020bzy}. 

The expected and observed limits at 95\% \CL obtained from the likelihood fits are shown in Fig.~\ref{fig:EFT_limit} (\cmsRight).
These limits are consistent with and generally comparable to previous aQGC measurements reported by CMS~\cite{CMS:2026photonWW,CMS:2020ssj,CMS:2021fyo,CMS:2021zby,CMS:2023wgjj,CMS:2025vvjj}. 

Finally, we combine with the previous measurement from CMS in the $\Pell\Pell\Pellpr\Pellpr$ channel~\cite{CMS:2020fqz} to assess the significance of EW \ZZ production. The combination is obtained from a simultaneous likelihood fit to the two channels, which are almost completely independent.
We tested correlating the signal theoretical uncertainties with those of the $\Pell\Pell\PGn\PGn$ analysis, and the impact is negligible: both channels are statistically limited with theoretical uncertainties constituting the dominant contribution among the systematic uncertainties, and the SR and CRs of the two analyses do not overlap.
We find an observed (expected) significance for EW production of \ZZ of 5.0 (4.5) standard deviations. The observed and expected significances of the individual channels and their combination are summarized in Table~\ref{tab:significance}.

\begin{table}[htbp]
    \topcaption{Observed and expected significances of EW \ZZ production in the $\Pell\Pell\PGn\PGn$ and $\Pell\Pell\Pellpr\Pellpr$ channels, and their combination.}
    \label{tab:significance}
    \centering
    \begin{scotch}{lcc}
    Channel & Expected & Observed \\
    \hline
    $\Pell\Pell\Pellpr\Pellpr$ & $3.5\sigma$ & $4.0\sigma$ \\
    $\Pell\Pell\PGn\PGn$ & $2.8\sigma$ & $3.1\sigma$ \\
    $\Pell\Pell\Pellpr\Pellpr + \Pell\Pell\PGn\PGn$ & $4.5\sigma$ & $5.0\sigma$ \\
\end{scotch}
\end{table}

Tabulated results are provided in the HEPData record for this analysis~\cite{hepdata}.

In summary, we report the first evidence for electroweak (EW) production of \PZ boson pairs in the final state $\PZ\PZ\text{jj}\to\Pell\Pell\PGn\PGn\text{jj}$, with a significance of 3.1 standard deviations.  
The measured fiducial cross section is in agreement with the standard model predictions. 
A search for anomalous quartic gauge couplings shows no evidence of beyond the standard model effects. Limits are set in terms of dimension-8 effective field theory operators.  
A combination with the previously reported result from the \ZZ decay channel with four charged leptons~\cite{CMS:2020fqz} yields an observed (expected) significance of 5.0 (4.5) standard deviations for the EW production of \PZ boson pairs.
With the addition of this result, the CMS experiment has observed EW production of all pairs of massive  gauge bosons:
\WW (both same-sign and opposite-sign), \WZ, and \ZZ.

\begin{acknowledgments}
We congratulate our colleagues in the CERN accelerator departments for the excellent performance of the LHC and thank the technical and administrative staffs at CERN and at other CMS institutes for their contributions to the success of the CMS effort. In addition, we gratefully acknowledge the computing centers and personnel of the Worldwide LHC Computing Grid and other centers for delivering so effectively the computing infrastructure essential to our analyses. Finally, we acknowledge the enduring support for the construction and operation of the LHC, the CMS detector, and the supporting computing infrastructure provided by the following funding agencies: SC (Armenia), BMBWF and FWF (Austria); FNRS and FWO (Belgium); CNPq, CAPES, FAPERJ, FAPERGS, and FAPESP (Brazil); MES and BNSF (Bulgaria); CERN; CAS, MoST, and NSFC (China); MINCIENCIAS (Colombia); MSES and CSF (Croatia); RIF (Cyprus); SENESCYT (Ecuador); ERC PRG and PSG, TARISTU24-TK10 and MoER TK202 (Estonia); Academy of Finland, MEC, and HIP (Finland); CEA and CNRS/IN2P3 (France); SRNSF (Georgia); BMFTR, DFG, and HGF (Germany); GSRI (Greece); MATE and NKFIH (Hungary); DAE and DST (India); IPM (Iran); SFI (Ireland); INFN (Italy); MSIT and NRF (Republic of Korea); MES (Latvia); LMTLT (Lithuania); MOE and UM (Malaysia); BUAP, CINVESTAV, CONACYT, LNS, SEP, and UASLP-FAI (Mexico); MOS (Montenegro); MBIE (New Zealand); PAEC (Pakistan); MSHE, NSC, and NAWA (Poland); FCT (Portugal); MESTD (Serbia); MICIU/AEI and PCTI (Spain); MOSTR (Sri Lanka); Swiss Funding Agencies (Switzerland); MST (Taipei); MHESI (Thailand); TUBITAK and TENMAK (T\"{u}rkiye); NASU (Ukraine); STFC (United Kingdom); DOE and NSF (USA).
\end{acknowledgments}\section*{Data availability} Release and preservation of data used by the CMS Collaboration as the basis for publications is guided by the  \href{https://doi.org/10.7483/OPENDATA.CMS.1BNU.8V1W}{CMS data preservation, re-use and open access policy}.

\bibliography{auto_generated} 

\appendix
\ifthenelse{\boolean{cms@external}}{\section{}}{\section{End Matter}\label{endmatter}}

\subsection{Single photon control region}
\label{endmatter:photonCR}

The method used to estimate the DY background uses the similarity between $\PGg+\text{jets}$ and $\PZ+\text{jets}$ events in boson kinematics and jet activities.  
In both processes, \ptmiss arises only from instrumental sources rather than from neutrinos.
Events in the single-photon CR are required to pass single-photon triggers with the lowest \pt threshold being 50\GeV, and further requirements on the reconstructed photon are applied, including $\pt(\PGg)>60\GeV$ and $\abs{\eta(\PGg)}<2.5$. 
It is important to subtract processes other than prompt photon production from the single-photon data sample, especially those that have genuine \ptmiss from neutrinos. 
The most important such processes are $\PZ\PGg+\text{jets}\to\PGn\PGn\PGg+\text{jets}$, $\PW+\text{jets} \to \Pe\PGn+\text{jets}$ where the electron is misidentified as a photon, and $\PW\PGg+\text{jets}\to \Pell\PGn\PGg+\text{jets}$ where the lepton is not detected. The first two processes are estimated from data and the last one is estimated from MC. 
The single-photon sample is then divided into two regions, $\ptmiss<120\GeV$ and $\ptmiss>120\GeV$.
In the lower-\ptmiss region, the photon sample is reweighted according to the number of reconstructed vertices, the boson \pt, and the boson $\eta$ to match a similar DY sample where the \PZ boson decays to electrons or muons. 
The weights are applied to the high-\ptmiss region of the single-photon CR and this gives the estimate of the DY background. 
The single-photon data alone cannot represent the kinematic distributions of the leptons from the \PZ boson decay.  Therefore, 
as a final step, we reweight the DY MC events to reproduce the \ptmiss and the boson \pt distributions observed in data, and this reweighted MC sample is used as the DY background estimate in all subsequent steps of the analysis. 

\subsection{Uncertainty estimation}
\label{endmatter:Uncertainty}
Systematic uncertainties are represented by nuisance parameters $\theta$ with probability density functions~\cite{CMS-NOTE-2011-005} centered on their nominal values $\hat{\theta}$. 
Correlations among systematic uncertainties in different categories are taken into account. The total likelihood is defined as the product of the likelihoods of the individual bins and the probability density functions for the nuisance parameters.
Uncertainties that affect only the overall normalization (\eg, the integrated luminosity) are modeled with log-normal probability density functions, while those that also affect the distributions of relevant kinematic observables (\eg, the JES) are treated as shape uncertainties.
Impacts are evaluated by performing the full analysis for not only the central value of the relevant parameter but also with the central value shifted up and down by one standard deviation. 

Discrepancies in the lepton reconstruction and identification efficiencies between data and simulation are 
corrected by applying data-to-simulation scale factors to all simulated samples. These scale factors are measured
using $\PZ/\PGgst \to \Pellp\Pellm$ events in the \PZ peak region~\cite{CMS:2018rym,CMS:2024ele} that are recorded with a set of triggers without lepton requirements.
The impact of the lepton momentum scale uncertainty is computed by varying the lepton momentum by 1\%.
The uncertainty in the JES and JER calibration affects the uncertainties in kinematic properties of jets and the \ptmiss.
In order to account for the systematic uncertainty of the JER smearing procedure, 
the resolution scale factors also vary up and down within their respective uncertainties.

Uncertainties in individual components used to build \ptmiss are propagated to the uncertainties in the \ptmiss distribution. The components include JES and JER uncertainties for jets, muon energy corrections, electron and photon energy scale uncertainties, and unclustered energy, comprising soft particles not clustered into jets, leptons or photons.
In this analysis, \PQb tagging is used only to reject events with candidate \PQb jets in the final state, and the uncertainties on the \PQb tagging efficiencies are included in the veto uncertainty. 

The uncertainty in the integrated luminosity is implemented with separate nuisance parameters accounting for year-dependent and partially correlated components across the 2016--2018 data-taking periods~\cite{CMS-PAS-LUM-20-001}, with the overall uncertainty of 0.73\%. 
Simulated samples are reweighted to reproduce the pileup conditions observed in the data. To compute the uncertainty related to pileup reweighting procedure, we alter the pileup distribution corresponding to the upward and downward $\Pp\Pp$ inelastic cross section variations of 5\%.  
The variation of the final yields induced by this procedure is less than 1\% for processes estimated from simulation.

Theoretical uncertainties in the normalization of the \ZZ, \WZ and signal processes are derived from variations of the QCD renormalization and factorization scales, 
$\alpS$, and PDFs~\cite{MCFM,Lai:2010vv,Martin:2009iq,Ball:2011mu,Butterworth:2015oua}. The scale uncertainties are taken as the largest deviation from the nominal prediction among independent variations of the renormalization and factorization scales by factors of two, excluding the two extreme variations.
The PDF and $\alpS$ uncertainties for signal and background processes are estimated from the standard deviation 
of weights from the replicas provided in the NNPDF3.1 PDF sets~\cite{nnpdf,NNPDF:2017mvq}. The samples for the signals and leading backgrounds are generated at NLO in QCD.

For the \ZZ process, the cross section is corrected to NNLO in QCD, but the assumed uncertainty reflects the NLO generator precision.   
The final distribution for the variable of interest is then reweighted bin-by-bin using the PDF4LHC prescription~\cite{Butterworth:2015oua}. The main contributions to the systematic uncertainty are renormalization and factorization scales, and PDFs, each contributing about 12\%. 

Another source of uncertainty in the estimation of the QCD $\text{diboson}+\text{jets}$ background comes from the missing higher-order EW terms in the event generation of the diboson processes. 
Terms purely of higher order in $\alpha_{\mathrm{EW}}$ are expected to be negligible~\cite{CMS:2020ulv}.
A significant uncertainty, instead, stems from missing mixed QCD--EW corrections of order $\alpS\alpha_{\mathrm{EW}}$.
For \ZZ events the uncertainty from such missing QCD--EW combined corrections is estimated from the product of the NLO QCD corrections ($1+\delta_{\mathrm{QCD}}$) 
and NLO EW corrections ($1+\delta_{\mathrm{EW}}$), as $\delta_{\mathrm{QCD}}\delta_{\mathrm{EW}}$. The NNLO QCD correction increases the predicted cross section by about 15\%, whereas the NLO EW correction reduces it by 5--25\% over the relevant \PZ boson \pt range, resulting in a net correction of approximately +10 to -10\%.
For the \WZ process, the EW correction is small~\cite{Bierweiler:2013dja,Gieseke:2014gka,Baglio:113005} and uncertainties are estimated from considering QCD--EW cross terms~\cite{Alwall:2014hca,Manohar:2016nzj,Frixione:2014qaa,Frixione:2015zaa}.\cleardoublepage \section{The CMS Collaboration \label{app:collab}}\begin{sloppypar}\hyphenpenalty=5000\widowpenalty=500\clubpenalty=5000\cmsinstitute{Yerevan Physics Institute, Yerevan, Armenia}
{\tolerance=6000
A.~Gevorgyan\cmsorcid{0000-0003-2751-9489}, A.~Hayrapetyan, V.~Makarenko\cmsorcid{0000-0002-8406-8605}, A.~Tumasyan\cmsAuthorMark{1}\cmsorcid{0009-0000-0684-6742}
\par}
\cmsinstitute{Marietta Blau Institute for Particle Physics, Vienna, Austria}
{\tolerance=6000
W.~Adam\cmsorcid{0000-0001-9099-4341}, L.~Benato\cmsorcid{0000-0001-5135-7489}, T.~Bergauer\cmsorcid{0000-0002-5786-0293}, M.~Dragicevic\cmsorcid{0000-0003-1967-6783}, S.~Gundacker\cmsorcid{0000-0003-2087-3266}, A.K.~Guven\cmsorcid{0009-0004-5670-5138}, P.S.~Hussain\cmsorcid{0000-0002-4825-5278}, M.~Jeitler\cmsAuthorMark{2}\cmsorcid{0000-0002-5141-9560}, N.~Krammer\cmsorcid{0000-0002-0548-0985}, A.~Li\cmsorcid{0000-0002-4547-116X}, D.~Liko\cmsorcid{0000-0002-3380-473X}, M.~Matthewman, J.~Schieck\cmsAuthorMark{2}\cmsorcid{0000-0002-1058-8093}, R.~Sch\"{o}fbeck\cmsAuthorMark{2}\cmsorcid{0000-0002-2332-8784}, M.~Shooshtari\cmsorcid{0009-0004-8882-4887}, M.~Sonawane\cmsorcid{0000-0003-0510-7010}, N.~Van~Den~Bossche\cmsorcid{0000-0003-2973-4991}, W.~Waltenberger\cmsorcid{0000-0002-6215-7228}, C.E.~Wulz\cmsAuthorMark{2}\cmsorcid{0000-0001-9226-5812}
% ...
\par}
$^{1}$Also at Yerevan State University, Yerevan, Armenia\\
$^{2}$Also at Technische Universit\"{a}t Wien, Vienna, Austria\\
% ...
\end{sloppypar}
%%% END EDITABLE REGION %%%
% skeleton_end
\end{document}